\documentclass{article}
\usepackage{authblk}
\usepackage[utf8]{inputenc}
\usepackage[T1]{fontenc}
\usepackage[USenglish]{babel}      
\usepackage{amsmath, amssymb}        
\usepackage{graphicx}                
\usepackage{hyperref} 
\usepackage{physics}					 
\usepackage{siunitx}                
\usepackage{textcomp}
\usepackage{gensymb} 				
\usepackage{url}					
\usepackage{booktabs}              
\usepackage{tabularx}
\title{Two-stage, low noise quantum frequency conversion of single photons from silicon-vacancy centers in diamond to the telecom C-band}
\author{Marlon Sch{\"a}fer}
\author{Benjamin Kambs}
\author{Dennis Herrmann}
\author{Tobias Bauer}
\author{Christoph Becher}
\affil{Fachrichtung Physik, Universit{\"a}t des Saarlandes, Campus E2 6, 66123 Saarbr{\"u}cken, Germany}

\begin{document}
\maketitle
\begin{abstract}
    The silicon-vacancy center in diamond holds great promise as a qubit for quantum communication networks. However, since the optical transitions are located within the visible red spectral region, quantum frequency conversion to low-loss telecommunication wavelengths becomes a necessity for its use in long-range, fiber-linked networks. This work presents a highly efficient, low-noise quantum frequency conversion device for photons emitted by a silicon-vacancy (SiV) center in diamond to the telecom C-band. By using a two-stage difference-frequency mixing scheme SPDC noise is circumvented and Raman noise is minimized, resulting in a very low noise rate of $10.4 \pm 0.7$ photons per second as well as an overall device efficiency of \SI{35.6}{\%}. By converting single photons from SiV centers we demonstrate the preservation of photon statistics upon conversion.
\end{abstract}

\section{Introduction}%
The vast majority of systems suitable as a quantum emitter or memory for quantum communications feature optical transitions in the visible and near infrared spectral region, experiencing strong absorption losses in optical fibers. For this reason, quantum frequency conversion (QFC) into low-loss telecom bands in combination with advanced concepts of quantum communication such as quantum repeaters \cite{San2011,azuma2022quantum} is the key enabling technology for long-range fiber-based quantum networks.  Using this technology, important primitives of quantum network elements were realized recently, e.g. a telecom-wavelength quantum repeater node with trapped ions \cite{Krut23}, entanglement of remote rubidium atom quantum memories via telecom photons \cite{vanLeent2021EntanglingSA,Luo2022} and two-photon interference from independent NV centers in diamond \cite{Stolk2022}, representing an advanced hardware platform for quantum networks \cite{Hermans2021QubitTB}.\\
Among the various hardware platforms for quantum communication the silicon-vacancy (SiV) center in diamond stands out due to a number of favorable properties \cite{becker2017}. In particular, the long spin coherence time \cite{sukachev2017}, Fourier-limited linewidths \cite{sip2014}, and excellent coupling to nanophotonic resonators with high cooperativity \cite{nguyen2019, Evans2018} enabled the demonstration of essential elements of quantum repeaters \cite{Nguyen_Oct2019,bhaskar2020, Knall2022}.
However, quantum frequency conversion of SiV photons into the telecom C-Band is particularly demanding. Direct conversion schemes using a \SI{1409}{\nm} pump beam to reach the target wavelength of \SI{1550}{\nm} suffer from strong pump-induced noise caused by Raman scattering and spontaneous parametric down-conversion (SPDC), prohibitive of reaching the single photon conversion regime \cite{Zaske2011}. Similar constraints hold for the direct conversion of photons from NV centers in diamond, typically pursued employing pump light at 1064 nm \cite{Dr2018, geus2022, mann2023lownoise}. \\ 
We here present efficient and low-noise QFC of single SiV photons following a two-stage conversion scheme. Periodically poled lithium niobate (PPLN) waveguides are used to first transduce the SiV photons to an intermediate wavelength, followed by a subsequent second conversion to the target telecom wavelength. For the two-stage difference frequency generation, in contrast to direct conversion, the chosen pump wavelength at \SI{2812.6}{\nm} is far above the target wavelength. Thereby, we circumvent SPDC noise and minimize Raman noise. 
\\%
This two-stage QFC technique was proposed by Pelc et al. \cite{pelc2012} and first implemented by Esfandyarpour et al. \cite{Esfan2018} in a QFC device down-converting \SI{650}{\nm} photons to \SI{1590}{\nm} in two cascaded  waveguides integrated on a chip. There, an overall device efficiency of 36 \% was achieved, however, the converted signal was not coupled into a fiber, but measured in free space. A further implementation of two-stage QFC demonstrated conversion of photons from a trapped Ba$^+$ ion at  at \SI{493}{\nm} to the telecom C-band, albeit with efficiencies of a few percent \cite{Han2021}, and, recently, to the telecom O-band with low noise and 11 \% overall efficiency \cite{Saha2023}.  \\
\section{Two-stage QFC device}%
\begin{figure}
    \includegraphics[width=\textwidth]{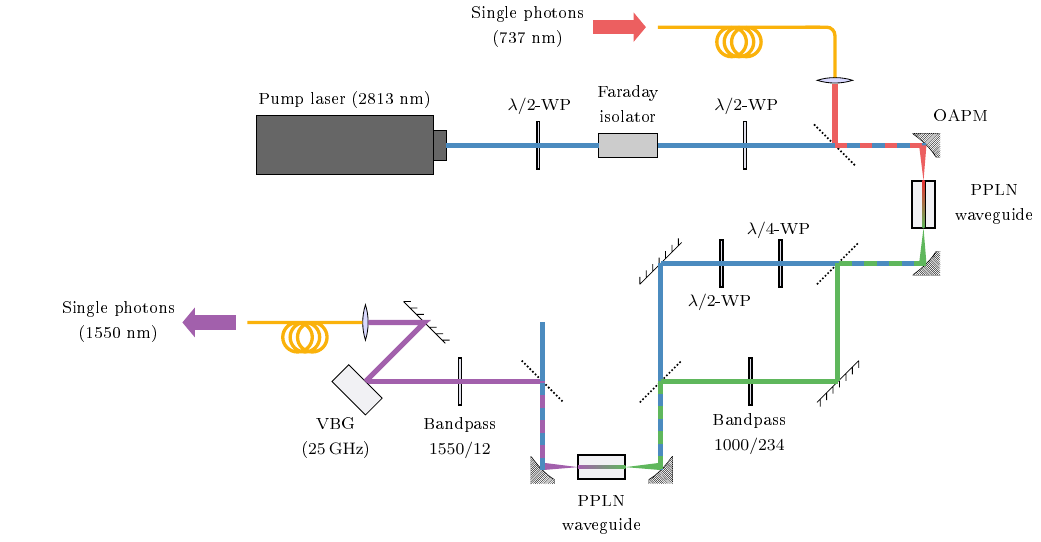}
    \caption{Schematic representation of the two-stage quantum frequency conversion set-up. Single photons resonant to SiV centers ($\SI{737}{\nm}$, red) are mixed with a strong pump beam ($\SI{2813}{\nm}$, blue) and down-converted in two separated periodically poled lithium niobate (PPLN) waveguides. In the first waveguide conversion to an intermediate wavelength ($\SI{999}{\nm}$, green) takes place, which is then subsequently transduced to the target telecom wavelength ($\SI{1549}{\nm}$, purple). By using \SI{90}{\degree} off-axis parabolic mirrors (OAPM) signal and pump wavelength are simultaneously coupled to the PPLN waveguide without chromatic aberration. Waveplates (WP) are used to manipulate the polarization of the pump light in the two waveguides and thus effectively control the pump power contributing to the DFG process in both waveguides independently. Broadband filtering with a bandpass and narrowband filtering with a Volume Bragg Grating (VBG) of \SI{25}{\GHz} bandwidth clears the converted signal of pump-induced noise photons before coupling it into the output fiber.}
    \label{fig:Aufbau}
\end{figure}

The quantum frequency conversion set-up transduces photons from a wavelength of \SI{737.1}{\nm} via \SI{998.9}{\nm} to \SI{1549.0}{\nm} in a two-stage, cascaded difference frequency generation process using two separate nonlinear crystals and the same pump wavelength of \SI{2812.6}{\nm} for both conversion stages. \\
A Cr$^{2+}$:ZnSe laser (\textit{IPG Photonics}) is used to generate a high intensity, single mode, single frequency pump beam. The laser is tunable in a range of \SI{2808}{\nm} to \SI{2820}{\nm}, however, due to absorption bands in air \cite{Ger1970} not all pump wavelengths are equally suitable. \\  
The experimental set-up is schematically depicted in Figure \ref{fig:Aufbau}. 
It consists of two separate PPLN waveguides (\textit{NTT Electronics}), where the single photons and the high-intensity pump beam are coupled in simultaneously. The waveguides are temperature controlled via a Peltier element. 
The individual beams of different wavelengths are combined and split by means of dichroic mirrors (\textit{Layertec}). In order to be able to set the pump power for both stages independently, we exploit the fact that only the fraction in s-polarization is relevant for the type-0 conversion process employed here. The ideal pump power for each conversion step is thus adjusted by rotating the linear polarization in front of the waveguides. In contrast to Esfandyarpour et al. \cite{Esfan2018} we do not use two waveguides integrated on the same chip, but two spatially separate waveguide chips instead. Thanks to this approach, the temperatures of the waveguides can be set independently, and thus the phase-matching temperatures for both DFG processes do not have to match. Moreover we can control the pump power contributing to the conversion in both stages independently by manipulating its polarization. \\ 
We use 90$^\circ$ off-axis parabolic mirrors (OAPM) to simultaneously couple the beams with different wavelengths into the waveguides, thereby avoiding chromatic aberration. For the same reason parabolic mirrors have already been used before in a set-up for polarization preserving frequency conversion by Krutyanskiy et al. \cite{krutyanskiy2017}.
After the first conversion stage, the transmitted pump light and the single photons are separated again with dichroic mirrors to allow for spectral filtering of the intermediate wavelength and polarization manipulation of the pump. As lithium niobate shows birefringence, the pump beam may be elliptically polarized after passing the first conversion crystal. For this reason, a quarter-wave plate is needed in addition to a half-wave-plate in order to to obtain an s-polarized pump beam for the second conversion stage. Noise photons photons induced in the first conversion stage are removed using a bandpass filter (\textit{Semrock}; bandwidth: \SI{234}{\nm}). 
Subsequently, the two wavelengths are overlapped again and coupled into the second waveguide, where conversion to \SI{1549}{\nm} takes place. 
A final dichroic mirror is used to separate the photons converted to the telecom C-band from the pump light. Before coupling into a single-mode fiber, a bandpass filter (\textit{Thorlabs}) with a bandwidth of \SI{12}{\nm} as well as a volume Bragg grating (\textit{Optigrate}, \SI{25}{\GHz} FWHM) clean the converted signal from noise photons. 
\\
\section{Performance of the QFC device}%
\begin{figure}
    \centering
    \includegraphics[width=\textwidth]{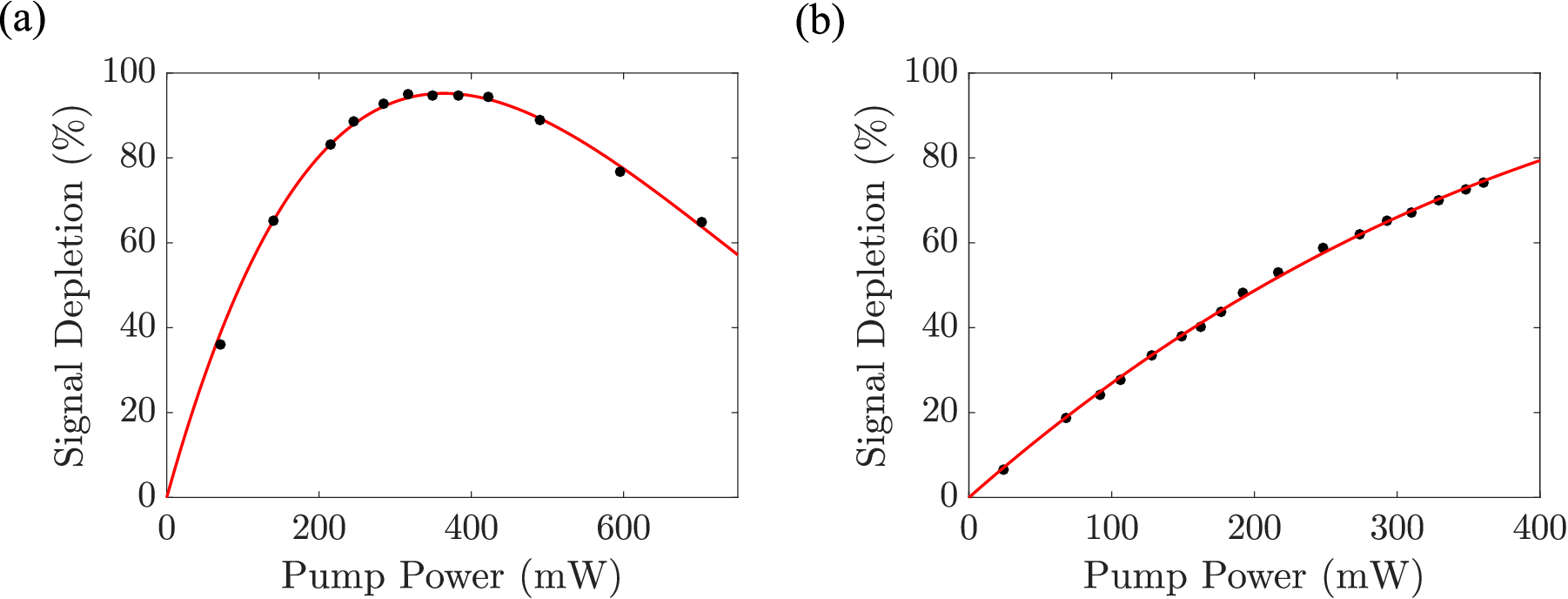}
    \caption{Signal depletion of the (a) first and (b) second conversion stage, the latter being limited by the available power of the pump beam. The data was fitted (red line) according to $\eta_\text{int} (P_\text{pump}) = \eta_\text{max}\sin^2(\sqrt{\kappa_\text{norm}P_\text{pump}}L)$, where $L=\SI{4}{\cm}$ is the length of the PPLN waveguide and the fit parameter $\eta_\text{max}$ denotes the maximum internal efficiency. }
    \label{fig:SignalDepletion}
\end{figure}

The conversion efficiency of the device was measured using a Ti:Sa laser tuned in resonance with the zero-phonon line of SiV centers at \SI{737.12}{\nm}. Here, we achieve an overall external device efficency of $\SI{35.6}{\%} \pm \SI{0.1}{\%}$, which is the conversion efficiency including all filtering and coupling losses, including coupling from and to single mode fibers. The influence of the different contributions to the external efficiency is detailed in the following.  \\
\begin{table}
	\centering
	\caption{\label{tab:extEff} The individual contributions to the measured external device efficiency of $\SI{35.6}{\%} \pm \SI{0.1}{\%}$. The transmission of the waveguides also includes the coupling efficiency. The measurement uncertainty, results from fluctuations in pump power and signal power during measurement.\\}
    \begin{tabular}{lc}
        \toprule%
        Contribution to external efficiency &  \% \\%
        \midrule%
        Transmission waveguide stage 1 	& $82.2 \pm 0.3$ \\%
        Signal depletion stage 1	&  $96.4 \pm 0.1$ \\%
        Bandpass 1000/234           &  $96.5 \pm 0.3$ \\%
        Transmission waveguide stage 2 	& $84.3 \pm 0.2$ \\%
        Signal depletion stage 2	&  $75.8 \pm 0.1$ \\%
        Bandpass 1550/12           &  $98.0 \pm 0.9$ \\%
        Volume Bragg grating        &  $97.6 \pm 0.5$ \\%
        Coupling efficiency in telecom fiber	& $83.0 \pm 0.4$ \\%
        Reflection/Transmission other optical components	&	$88.6 \pm 1.7$ \\%
        \midrule%
        $\Sigma$ 	& $34.5 \pm 0.8$ \\%
        \bottomrule%
    \end{tabular}%
\end{table}%
Internal efficiencies are $\SI{96.4}{\%} \pm \SI{0.1}{\%}$ and $\SI{75.8}{\%} \pm \SI{0.1}{\%}$ \% for the first respectively second conversion stage, resulting in an overall internal efficiency of $\SI{73.1}{\%} \pm \SI{0.1}{\%}$. Internal efficiencies were determined by measuring signal depletion as a function of pump power for both stages separately, which is shown in Figure \ref{fig:SignalDepletion}. 
As can be seen in Figure \ref{fig:SignalDepletion}b, the internal efficiency of the second conversion stage from \SI{999}{\nm} to telecom wavelength is limited by the amount of available pump power. The strong loss of pump power in the setup is significantly favored by the different absorption bands present at wavelengths around \SI{2.8}{\um}. To begin with, strong water absorption bands exist in the spectral region of the pump wavelength, leading to humidity dependent absorption in air \cite{Ger1970}. This being said, the wavelength of \SI{2812.6}{\nm} was chosen to minimize absorption in air, resulting in a loss of about \SI{10}{\%} up to the second waveguide. A second loss channel is given by absorption losses in the fused silica substrate of the dichroic mirrors. For fused silica it is well known that Hydroxy (OH) groups lead to a strong absorption band in the spectral region around \SI{2750}{\nm} \cite{Wal1978, Kum1981, Tom2001}. In the fused silica used here, this absorption band was significantly weakened due to a reduced OH content. Still, for each dichroic mirror an absorption of about 10 \% is measured, which is in agreement with the specifications by the vendor. Finally, also the PPLN waveguides are found to be a significant loss channel. For LiNbO$_3$ doped with 5 mol \% MgO and s-polarized light, absorption coefficients between \SI{0.08}{\cm^{-1}} \cite{Schwes2010} and \SI{0.088}{\cm^{-1}} \cite{Schwes2011} for the absoprtion maximum at \SI{2826}{\nm} are found in the literature. For the pump beam we measured a transmission of 63 \% and 68 \% for the first and second waveguide, respectively. However, we cannot specify the fraction of these losses due to absorption, since we cannot distinguish between absorption and coupling losses. \\
Further contributions to the external device efficiency, i.e. losses at spectral filters, mirrors as well as the coupling efficiency to the output single mode fiber, were all measured separately. The results are listed in Table \ref{tab:extEff}. This way, we obtain a calculated external efficiency of $34.5 \% \pm 0.8 \%$. The slight deviation from the external efficiency measured directly can be explained by fluctuations in the internal efficiency due to fluctuations of the power and spectral and spatial mode profiles of the pump laser. \\ 
Conversion noise is measured by by setting the pump power to the optimal operating point and blocking the signal input while detecting the photon rate in the output fiber with superconducting nanowire single-photon detectors (SNSPD, \textit{Single Quantum}). Integrating over 15 minutes yielded $17.4 \pm 0.2 \, \text{cps}$, of which $12.1 \pm 0.2 \, \text{cps}$ can be assigned to dark counts of the detectors. Substracting the dark counts and correcting by the detection efficiency of the SNSPD (72 \%) and transmission losses to the detector (89 \%) we get a noise photon rate induced by the pump during conversion of only $10.4 \pm 0.7$ photons per second or, correspondingly, 0.4 photons per second per gigahertz filter bandwidth. These few remaining noise photons can most probably be attributed to residual Raman scattering. Although the spectral gap between pump (\SI{2812.6}{\nm}) and target (\SI{1550}{\nm}) wavelengths is large (\SI{2900}{\cm^{-1}}), our assumption is supported by the measurement of a non-Lorentzian pedestal for high frequency shifts in the Raman spectrum of LiNbO$_3$ by Pelc et al. \cite{Pelc2011_2} showing the presence of Raman scattered photons at frequency shifts of $1600\,\text{cm}^{-1}$ and above. By comparison, for the two-stage QFC set-up of Esfandyarpour et al. \cite{Esfan2018} noise was estimated as $1.5\,\text{photons/s/GHz}$ from measured Raman spectra (spectral gap: \SI{1740}{\cm^{-1}}). The results are in good agreement as for the converter presented here a lower Raman noise is expected due the larger spectral gap between pump and target wavelength.
\section{Single photon conversion}%
\begin{figure}%
    \includegraphics[width=\textwidth]{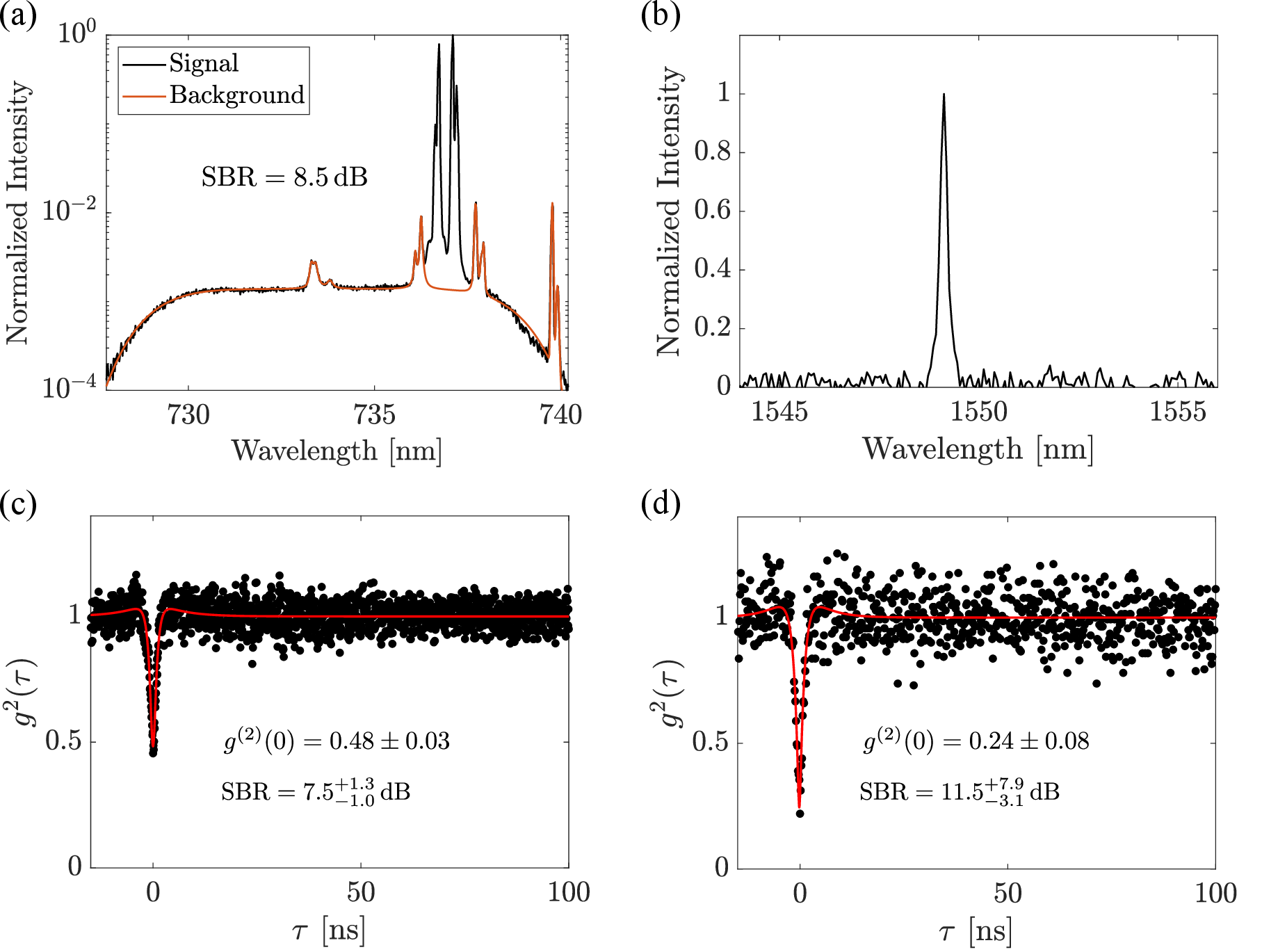}%
    \caption{(a) Spectrum of the SiV center used for conversion in semilogarithmic representation. A bandpass filter with a central wavelength of 735 nm and a linewidth of 11 nm is used as the detection filter. In orange the background flourescence from the sample is fitted with a super-gaussian and a sum of lorentz peaks. The signal-to-background ratio (SBR) is obtained by dividing the background subtracted signal and the background. (b) Spectrum of the single SiV photons after conversion. Due to the acceptance bandwith of 77(4) GHz only the C-transition of the SiV center is converted. The linewidth of the converted spectrum is limited by the spectrometer's resolution of about 20 GHz. (c) $g^{(2)}$-measurement of the unconverted photons. (d) $g^{(2)}$-measurement of the converted photons. The signal-to-background ratio is extracted by a fitting routine that uses the jitter of the detection unit as a fixed parameter.}%
    \label{fig:SinglePhotonsQFC}%
\end{figure}

To demonstrate the low-noise performance of the two-stage QFC we here prove the preservation of single photon statistics by measuring the second order auto-correlation function $g^{(2)}(\tau)$ before and after conversion. It has to be noted that the experiment was performed with an earlier version of the quantum frequency converter at that time achieving \SI{29}{\%} external efficiency and a higher noise photon rate of about $500\, \text{cps}$. The latter is due to the fact that the Volume Bragg Grating had not yet been integrated in the device and only a bandpass with \SI{12}{\nm} bandwith was used as a filter in front of the output fiber. Despite the higher noise count rate of the converter preservation of the single photon character is unambiguously shown.\\
Single photons were created from a sample consisting of diamond nanopillars containing SiV centers. The sample was cooled down in a helium flow cryostat to about \SI{10}{K}. Under non-resonant excitation at \SI{532}{\nm}, we obtained a photon rate of about $\SI{5.5e5}{}\,\text{cps}$ collected into a single-mode fiber. The emission spectrum shown in Figure \hyperref[fig:SinglePhotonsQFC]{3a} reveals the characteristic four line fine structure of the SiV center. \\
Photon correlation measurements are performed using Hanbury-Brown Twiss (HBT) setups with avalanche photodiodes (APD, \textit{Excelitas Technologies}) for photons at 737 nm and 
SNSPDs (\textit{Single Quantum}) for telecom photons. 
For the unconverted single photons a nonlinear least squares fit using a Levenberg-Marquardt algorithm yields a dip of the $g^{(2)}(\tau)$ function to $g^{(2)}(0) = 0.48 \pm 0.03$, see Figure \hyperref[fig:SinglePhotonsQFC]{3c}. The non-vanishing value of the $g^{(2)}$-function for $\tau=0$ can be fully accounted for by the jitter of the detectors and the background fluorescence of the sample. Using the signal-to-background (SBR) ratio as free fit parameter and setting the jitter to the \SI{550}{\ps} specified for of the APDs, the fit result indicates a SBR of \SI{7.5}{\dB} with the \SI{95}{\%} confidence interval ranging from \SI{6.5}{\dB} to \SI{8.8}{\dB}. This is in agreement with the value of \SI{8.5}{\dB} calculated from the measured spectrum in Figure \hyperref[fig:SinglePhotonsQFC]{3a} by fitting the background fluorescence from the sample with a a super-Gaussian and a sum of Lorentzian peaks. \\
Due to the QFC device acceptance bandwidth of $77 \,\text{GHz}$ only one fine structure line of the SiV spectrum can be converted. Since of all four fine structure lines most of the intensity is emitted into the C-transition, the acceptance bandwidth of the converter was centered around the C-transition at \SI{737.12}{\nm} by tuning the temperatures of the nonlinear crystals. As expected, the spectrum of the converted photons in \hyperref[fig:SinglePhotonsQFC]{3b} shows a single peak at \SI{1549}{\nm}. After conversion to telecom wavelength the measurement of the $g^{(2)}(\tau)$-function (cf. Figure \hyperref[fig:SinglePhotonsQFC]{3d}) reveals a single photon purity of $g^{(2)}(0) = 0.24 \pm 0.08$ which is better than for the unconverted photons. There are two reasons for this: On the one hand, the SNSPD's jitter of \SI{310}{\ps} (FWHM) is significantly lower compared to the APDs that were used for the unconverted photons. Second, and despite the additional noise photons induced by the pump light, we in total get a higher signal-to-background ratio due to the spectral filtering of the SiV background by the \SI{77}{\GHz} acceptance bandwidth of the QFC device.
Within the \SI{77}{\GHz} acceptance bandwidth the background fluorescence is negligible compared to the signal. The signal-to-background ratio of the converted photons is thus determined by the pump-induced noise during conversion. In this experiment, of the $20\, \text{kcps}$ single photon rate detected by the SNSPDs, $500\, \text{cps}$ was due to conversion-induced noise, hence, yielding a signal to noise ratio of \SI{15}{\dB} for the converted photons. This estimation is consistent with the fit result of the $g^{(2)}$-function, yielding a SBR of \SI{11.5}{\dB} with the \SI{95}{\%} confidence interval ranging from \SI{8.4}{\dB} to \SI{18.4}{\dB}. \\
Note that the $g^{(2)}$-measurement of the converted photons looks much more noisy compared to the unconverted photons, since only about $20\, \text{kcps}$ of the $550\, \text{kcps}$ emitted by the SiV center are left after conversion. This can be explained as follows: First, the acceptance bandwidth of the converter implies that only the C-transition of the fine structure can be converted containing only about \SI{40}{\%} of the total emission by the SiV center. Second, the conversion of the SiV photons took place in a different laboratory than the excitation of the emitter and photon collection. Due to losses at optical fiber patches, the transmission of the interconnecting fiber is about \SI{61}{\%}.
Another loss factor is the \SI{29}{\%} device efficiency of the converter. Eventually, the detection efficiency of the SNSPDs used in this experiment is \SI{35}{\%}, which is lower than the \SI{60}{\%} efficiency of the APDs. Combining all these factors yields for the converted photon rate: $550\, \text{kcps} \cdot 0.4 \cdot 0.61 \cdot 0.29 \cdot \frac{0.35}{0.6} = 23\, \text{kcps}$, which corresponds very well to the count rate during the $g^{(2)}$-measurement.
\section{Conclusion and outlook}%
In conclusion we have shown a very low-noise yet highly efficient quantum frequency conversion device for converting photons resonant to silicon-vacancy centers in diamond to the telecom C-band. We used a two-stage conversion scheme, where photons are first converted to an intermediate wavelength followed by a subsequent conversion to the target wavelength. In this way, we succeeded in achieving a very low conversion induced noise rate of $10.4 \pm 0.7$ photons per second or, taking into account the \SI{25}{\GHz} filter bandwidth, 0.4 photons per second per Gigahertz filter bandwidth. This small noise level should be compared to the direct, one-stage conversion of attenuated laser pulses at \SI{738}{\nm} to the telecom C-band which resulted in a noise rate of about 400'000 photons per second in a \SI{95}{\GHz} filter bandwidth \cite{Zaske2011}. Even though it was possible to reduce the noise rate by additional temporal filtering, a minimum noise level of 2000 photons per second or, equivalently, 21 photons per second per Gigahertz was achieved. This in addition required a reduction of the pump power and thus the internal efficiency to about 60 \%. The low noise level demonstrated here also favorably compares to other QFC schemes at shorter wavelengths \cite{Dr2018, Esfan2018, Han2021, geus2022, Saha2023, mann2023lownoise}.   \\
The overall device efficiency of \SI{35.6}{\%} is mainly limited by the pump power being absorbed in the optical components of the device, since the pump wavelength is in the range of the absorption bands of air, fused silica and lithium niobate. As a result, the remaining pumping power in the second conversion stage is not sufficient to achieve maximum conversion efficiency, leading to an internal efficiency of only \SI{76}{\%} in the second stage. \\
By increasing the pump power in the second conversion stage, either by using a more powerful pump laser or components with lower absorption, an internal efficiency of the second stage of \SI{90}{\%} or more is expected. 
Assuming further that it is possible to achieve the same coupling efficiencies as have been achieved in \cite{Leent2020} for \SI{780}{\nm} photons, i.e. \SI{87.6}{\%} for fiber coupling and \SI{90.0}{\%} for waveguide coupling, an improvement of the overall device efficiency to about \SI{50}{\%} should be feasible for a two-stage SiV converter. \\
Finally, we successfully converted single photons emitted by a silicon-vacancy center in diamond and demonstrated the preservation of the single photon statistics upon conversion. Despite the conversion-induced noise photons, the single photon purity improved to a value of $g^{(2)}(0) = 0.24 \pm 0.08$ after conversion, which can be explained by the spectral filtering of the background fluorescence within the conversion process. The single photon conversion results presented here for an early version of the converter are potentially further improved by the better noise performance of the final device. \\
Looking ahead, the two-stage conversion design presented here can also be used to the convert single photons of other other single quantum emitters or qubit systems that suffer from similar high noise rates after direct, one-stage quantum frequency conversion, e.g. the nitrogen-vacancy \cite{Dr2018} and the tin-vacancy center \cite{Ikuta2014} in diamond, emitting at \SI{637}{\nm} and \SI{619}{\nm}, respectively.
\section*{Acknowledgment}%
We thank Johannes Görlitz for help with the single photon conversion experiments and David Lindler for helpful discussions.
We acknowledge support by the German Federal Ministry of Education and Research (Bundesministerium für Bildung und Forschung, BMBF) through projects HiFi (13N15926) and QR.X (16KISQ001K).
\bibliography{main}

\begin{thebibliography}{10}
\providecommand{\url}[1]{\texttt{#1}}
\providecommand{\urlprefix}{URL }

\bibitem{San2011}
N.~Sangouard, C.~Simon, H.~de~Riedmatten, N.~Gisin,
\newblock \emph{Rev. Mod. Phys.} \textbf{2011}, \emph{83} 33.

\bibitem{azuma2022quantum}
K.~Azuma, S.~E. Economou, D.~Elkouss, P.~Hilaire, L.~Jiang, H.-K. Lo,
  I.~Tzitrin,
\newblock Quantum repeaters: From quantum networks to the quantum internet,
  \textbf{2022}.

\bibitem{Krut23}
V.~Krutyanskiy, M.~Canteri, M.~Meraner, J.~Bate, V.~Krcmarsky, J.~Schupp,
  N.~Sangouard, B.~P. Lanyon,
\newblock \emph{Phys. Rev. Lett.} \textbf{2023}, \emph{130} 213601.

\bibitem{vanLeent2021EntanglingSA}
T.~van Leent, M.~Bock, F.~Fertig, R.~Garthoff, S.~Eppelt, Y.~Zhou, P.~Malik,
  M.~M. Seubert, T.~Bauer, W.~Rosenfeld, W.~Zhang, C.~Becher, H.~Weinfurter,
\newblock \emph{Nature} \textbf{2021}, \emph{607} 69 .

\bibitem{Luo2022}
X.-Y. Luo, Y.~Yu, J.-L. Liu, M.-Y. Zheng, C.-Y. Wang, B.~Wang, J.~Li, X.~Jiang,
  X.-P. Xie, Q.~Zhang, X.-H. Bao, J.-W. Pan,
\newblock \emph{Phys. Rev. Lett.} \textbf{2022}, \emph{129} 050503.

\bibitem{Stolk2022}
A.~Stolk, K.~van~der Enden, M.-C. Roehsner, A.~Teepe, S.~Faes, C.~Bradley,
  S.~Cadot, J.~van Rantwijk, I.~te~Raa, R.~Hagen, A.~Verlaan, J.~Biemond,
  A.~Khorev, R.~Vollmer, M.~Markham, A.~Edmonds, J.~Morits, T.~Taminiau, E.~van
  Zwet, R.~Hanson,
\newblock \emph{PRX Quantum} \textbf{2022}, \emph{3} 020359.

\bibitem{Hermans2021QubitTB}
S.~L.~N. Hermans, M.~Pompili, H.~K.~C. Beukers, S.~Baier, J.~Borregaard,
  R.~Hanson,
\newblock \emph{Nature} \textbf{2021}, \emph{605} 663 .

\bibitem{becker2017}
J.~N. Becker, C.~Becher,
\newblock \emph{physica status solidi (a)} \textbf{2017}, \emph{214}, 11
  1700586.

\bibitem{sukachev2017}
D.~D. Sukachev, A.~Sipahigil, C.~T. Nguyen, M.~K. Bhaskar, R.~E. Evans,
  F.~Jelezko, M.~D. Lukin,
\newblock \emph{Phys. Rev. Lett.} \textbf{2017}, \emph{119} 223602.

\bibitem{sip2014}
A.~Sipahigil, K.~D. Jahnke, L.~J. Rogers, T.~Teraji, J.~Isoya, A.~S. Zibrov,
  F.~Jelezko, M.~D. Lukin,
\newblock \emph{Phys. Rev. Lett.} \textbf{2014}, \emph{113} 113602.

\bibitem{nguyen2019}
C.~T. Nguyen, D.~D. Sukachev, M.~K. Bhaskar, B.~Machielse, D.~S. Levonian,
  E.~N. Knall, P.~Stroganov, R.~Riedinger, H.~Park,
  M.~Lon\ifmmode~\check{c}\else \v{c}\fi{}ar, M.~D. Lukin,
\newblock \emph{Phys. Rev. Lett.} \textbf{2019}, \emph{123} 183602.

\bibitem{Evans2018}
R.~E. Evans, M.~K. Bhaskar, D.~D. Sukachev, C.~T. Nguyen, A.~Sipahigil, M.~J.
  Burek, B.~Machielse, G.~H. Zhang, A.~S. Zibrov, E.~Bielejec, H.~Park,
  M.~Lončar, M.~D. Lukin,
\newblock \emph{Science} \textbf{2018}, \emph{362}, 6415 662.

\bibitem{Nguyen_Oct2019}
C.~T. Nguyen, D.~D. Sukachev, M.~K. Bhaskar, B.~Machielse, D.~S. Levonian,
  E.~N. Knall, P.~Stroganov, C.~Chia, M.~J. Burek, R.~Riedinger, H.~Park,
  M.~Lon\ifmmode~\check{c}\else \v{c}\fi{}ar, M.~D. Lukin,
\newblock \emph{Phys. Rev. B} \textbf{2019}, \emph{100} 165428.

\bibitem{bhaskar2020}
M.Bhaskar, R.~Riedinger, B.~Machielse, D.~Levonian, C.~Nguyen, E.~Knall,
  H.~Park, D.~Englund, M.~Loncar, D.~Sukachev, M.~Lukin,
\newblock \emph{Nature} \textbf{2020}, \emph{580} 60{\textendash}64.

\bibitem{Knall2022}
E.~N. Knall, C.~M. Knaut, R.~Bekenstein, D.~R. Assumpcao, P.~L. Stroganov,
  W.~Gong, Y.~Q. Huan, P.-J. Stas, B.~Machielse, M.~Chalupnik, D.~Levonian,
  A.~Suleymanzade, R.~Riedinger, H.~Park, M.~Lon\ifmmode~\check{c}\else
  \v{c}\fi{}ar, M.~K. Bhaskar, M.~D. Lukin,
\newblock \emph{Phys. Rev. Lett.} \textbf{2022}, \emph{129} 053603.

\bibitem{Zaske2011}
S.~Zaske, A.~Lenhard, C.~Becher,
\newblock \emph{Opt. Express} \textbf{2011}, \emph{19}, 13 12825.

\bibitem{Dr2018}
A.~Dr\'eau, A.~Tchebotareva, A.~E. Mahdaoui, C.~Bonato, R.~Hanson,
\newblock \emph{Phys. Rev. Applied} \textbf{2018}, \emph{9} 064031.

\bibitem{geus2022}
J.~F. Geus, F.~Elsen, S.~Nyga, B.~Jungbluth, H.-D. Hoffmann, C.~Haefner,
\newblock In P.~R. Hemmer, A.~L. Migdall, editors, \emph{Quantum Computing,
  Communication, and Simulation II}, volume 12015. International Society for
  Optics and Photonics, SPIE, \textbf{2022} 1201506.

\bibitem{mann2023lownoise}
F.~Mann, H.~M. Chrzanowski, F.~Gewers, M.~Placke, S.~Ramelow,
\newblock Low-noise quantum frequency conversion in a monolithic bulk ppktp
  cavity, \textbf{2023},
\newblock \textit{{a}rXiv preprint arXiv:2304.13459.}

\bibitem{pelc2012}
J.~S. Pelc,
\newblock Ph.D. thesis, Stanford University, \textbf{2012}.

\bibitem{Esfan2018}
V.~Esfandyarpour, C.~Langrock, M.~Fejer,
\newblock \emph{Opt. Lett.} \textbf{2018}, \emph{43}, 22 5655.

\bibitem{Han2021}
J.~Hannegan, U.~Saha, J.~D. Siverns, J.~Cassell, E.~Waks, Q.~Quraishi,
\newblock \emph{Applied Physics Letters} \textbf{2021}, \emph{119}, 8 084001.

\bibitem{Saha2023}
U.~Saha, J.~D. Siverns, J.~Hannegan, Q.~Quraishi, E.~Waks,
\newblock \emph{ACS Photonics} \textbf{0}, \emph{0}, 0 null.

\bibitem{Ger1970}
C.~Gerritsma, J.~Haanstra,
\newblock \emph{Infrared Physics} \textbf{1970}, \emph{10}, 2 79.

\bibitem{krutyanskiy2017}
V.~{Krutyanskiy}, M.~{Meraner}, J.~{Schupp}, B.~P. {Lanyon},
\newblock \emph{Applied Physics B: Lasers and Optics} \textbf{2017},
  \emph{123}, 9 228.

\bibitem{Wal1978}
G.~E. Walrafen, S.~R. Samanta,
\newblock \emph{The Journal of Chemical Physics} \textbf{1978}, \emph{69}, 1
  493.

\bibitem{Kum1981}
B.~Kumar, N.~Fernelius, J.~A. Detrio,
\newblock \emph{Journal of the American Ceramic Society} \textbf{1981},
  \emph{64}, 12 C‐178.

\bibitem{Tom2001}
M.~Tomozawa, D.-L. Kim, V.~Lou,
\newblock \emph{Journal of Non-Crystalline Solids} \textbf{2001}, \emph{296}, 1
  102.

\bibitem{Schwes2010}
J.~R. Schwesyg, C.~R. Phillips, K.~Ioakeimidi, M.~C.~C. Kajiyama, M.~Falk,
  D.~H. Jundt, K.~Buse, M.~M. Fejer,
\newblock \emph{Opt. Lett.} \textbf{2010}, \emph{35}, 7 1070.

\bibitem{Schwes2011}
J.~R. Schwesyg, A.~Markosyan, M.~Falk, M.~C.~C. Kajiyama, D.~H. Jundt, K.~Buse,
  M.~M. Fejer,
\newblock In \emph{Advances in Optical Materials}. Optical Society of America,
  \textbf{2011} AIThE3.

\bibitem{Pelc2011_2}
J.~S. Pelc, L.~Ma, C.~R. Phillips, Q.~Zhang, C.~Langrock, O.~Slattery, X.~Tang,
  M.~M. Fejer,
\newblock \emph{Opt. Express} \textbf{2011}, \emph{19}, 22 21445.

\bibitem{Leent2020}
T.~van Leent, M.~Bock, R.~Garthoff, K.~Redeker, W.~Zhang, T.~Bauer,
  W.~Rosenfeld, C.~Becher, H.~Weinfurter,
\newblock \emph{Phys. Rev. Lett.} \textbf{2020}, \emph{124} 010510.

\bibitem{Ikuta2014}
R.~Ikuta, T.~Kobayashi, S.~Yasui, S.~Miki, T.~Yamashita, H.~Terai, M.~Fujiwara,
  T.~Yamamoto, M.~Koashi, M.~Sasaki, Z.~Wang, N.~Imoto,
\newblock \emph{Opt. Express} \textbf{2014}, \emph{22}, 9 11205.

\end{thebibliography}
\pagestyle{empty}
\end{document}